\documentclass[12pt,letterpaper,superscriptaddress]{JHEP3}
\usepackage{amssymb,amsmath,amsfonts,amstext,graphics}
\usepackage{cite}
\usepackage{graphicx}
\usepackage{epstopdf}
\input epsf.tex

\newcommand{\gsim}{\lower.7ex\hbox{$\;\stackrel{\textstyle>}{\sim}\;$}}
\newcommand{\lsim}{\lower.7ex\hbox{$\;\stackrel{\textstyle<}{\sim}\;$}}
\def\GG{{\cal G}}

\def\LL{{\cal L}}
\def\OO{{\cal O}}
\def\PP{{\cal P}}

\def\eg{{\it e.g. }}

\newcommand{\TeV}{\,\mathrm{TeV}}

\newcommand{\half}{{\frac{1}{2}  }}

\newcommand{\identity}{{\rlap{1} \hskip 1.6pt \hbox{1}}}

\newcommand{\Tr}{{\text{ Tr }}}

\newcommand{\bef}{\begin{figure}[htbp]\begin{center}}
\newcommand{\eef}{\end{center}\end{figure}}

\newcommand{\nlsm}{{nl$\sigma$m}}

\title{
\begin{flushright}
\mbox{\normalsize SLAC-PUB-13865}
\end{flushright}
\vskip 15 pt
Collective Quartics from Simple Groups}
\author{Anson Hook and Jay G. Wacker \\
Theory Group,\\ 
Stanford Linear Accelerator Center,\\
Menlo Park, CA 94025}
 
\abstract{
This article classifies Little Higgs models that have collective quartic couplings.
There are two classes of collective quartics:  Special Cosets and Special Quartics.
After taking into account dangerous singlets, the smallest Special Coset models are $SU(5)/SO(5)$ and $SU(6)/Sp(6)$.  The smallest Special Quartic model is $SU(5)/SU(3)\times SU(2)\times U(1)$ and
has not previously been considered as a candidate Little Higgs model.
}

\begin{document}

\section{Introduction}

Little Higgs (LH) models are a leading candidate for extensions to the Standard Model.
The primary advance of Little Higgs models 
\cite{ArkaniHamed:2001nc, ArkaniHamed:2002pa,ArkaniHamed:2002qx,Chang:2003un,Katz:2003sn,Kaplan:2003uc,Kaplan:2004cr,Schmaltz:2004de} 
over older models where the Higgs is a Goldstone boson 
\cite{Georgi:1975tz,Kaplan:1983fs,Kaplan:1983sm,Dugan:1984hq, Georgi:1984af,Georgi:1984ef}
is the existence of an operator that gives a quartic coupling, but no mass term.
The origin of the independent quartic coupling arises from collective symmetry breaking -- where two separate symmetries treat the Higgs as a Goldstone boson.
If either of these symmetries is exact, then the Higgs is a massless Goldstone boson; however, when both symmetries are broken, the Higgs
can have a potential.  
Collective symmetry breaking guarantees that the one loop quadratic divergences in the theory renormalize operators that do not induce a mass for the Higgs boson.   In many theories ({\it e.g.}~the Littlest Higgs  \cite{ArkaniHamed:2002qy} or a Little Higgs from an Anti-Symmetric Condensate \cite{Low:2002ws}), two of the operators that had a one loop quadratic divergence, when taken together, induced a quartic coupling for the Higgs boson without inducing a mass at the same time.  
  This article explains the structure in the theory that enables these models to have operators with a quartic coupling without a mass: ``{\it collective quartics}.''

The existence of a quartic coupling independent of the mass term is critical because
without an independent quartic coupling, electroweak symmetry breaking must
arise from vacuum misalignment and frequently results in either a light Higgs boson or fine tuning \cite{Kaplan:1983fs, Banks:1984gj, Gregoire:2003kr}.

Simple group Little Higgs models are arguably the most elegant version of Little Higgs models.  
These models are cosets $G/H$ where $G$ is simple, \eg
 the Littlest Higgs ($SU(5)/SO(5)$) \cite{ArkaniHamed:2002qy}, a Little Higgs from an Anti-Symmetric Condensate ($SU(6)/Sp(6)$) \cite{Low:2002ws},
$SU(9)/SU(8)$ \cite{Skiba:2003yf}, and a Simple Custodially Symmetric Little Higgs ($SO(9)/SO(5)\times SO(4)$) \cite{Chang:2003zn}.
Additionally, there is the Intermediate Higgs ($SU(4)/Sp(4)$) which is not a proper Little Higgs model
but can soften the top quadratic divergence \cite{Katz:2005au}.  

Recently, Schmaltz and Thaler showed that the existence of a radiatively stable collective quartic coupling
places restrictions on the scalar field content of the cosets \cite{Schmaltz:2008vd}.  Specifically, they found that uncharged singlets that
participate in the collective quartics have tadpoles that reintroduce quadratic divergences to the
Higgs mass.   Schmaltz and Thaler left open the question of the smallest coset with a collective quartic 
and no dangerous tadpole.

This article classifies the existence of collective quartics without dangerous tadpoles from simple Little Higgs or simple Intermediate Higgs models that satisfy the following criteria.
The Little Higgs coset arises from the spontaneous breaking of $G$ to $H$.  In this article, $G$ is a simple group.   In the breaking from $G$ to $H$, the electroweak gauge symmetry is {\it not} broken.  
Electroweak generators need to be embedded inside of $H$. The Higgs arises as a pseudo-Goldstone boson (PGB) from this breaking.  
Little Higgs models frequently have an extended electroweak gauge sector that is broken when $G$ is broken to $H$; however, the analysis presented in this article does not require specifying the gauge symmetry.  At the cost of reintroducing the gauge quadratic divergences to the Higgs mass (but still softening the top's quadratic divergence), the TeV gauge symmetry can be the electroweak gauge group, as in the Intermediate Higgs.  This article will present a new model along the lines of the Intermediate Higgs that softens the quadratic divergences of the top and Higgs sector and has a radiatively safe quartic coupling without dangerous tadpoles.
The smallest possible model with a collective quartic is found to be $SU(5)/SU(3)\times SU(2)\times U(1)$ with 12 Goldstone bosons.

\subsection{Collective Quartics}

The existence of a collective quartic coupling
places restrictions upon the possible groups because it
is generated from a potential of the form
\begin{eqnarray}
\label{Eq: General LH Potential}
V(\Sigma) = 
\lambda_1 f^4 \Tr \PP_1 \Sigma \PP^({}'{}^)_1 \Sigma^\dagger
+\lambda_2 f^4 \Tr \PP_2\Sigma \PP^({}'{}^)_2 \Sigma^\dagger
\end{eqnarray}
where $\Sigma$ is the non-linear sigma model (\nlsm) field and $\PP_1$ and $\PP_2$ are projection operators that preserve a subgroup of $G$, $\GG_1$ and $\GG_2$, respectively.    
$\GG_1$ and $\GG_2$ are  nonlinearly realized subgroups that shift the Higgs boson and do not commute with $H$.
If either $\lambda_1$ or $\lambda_2$ vanish, then 
$\GG_1$ or $\GG_2$ becomes an exact symmetry and the
Higgs boson responsible for electroweak symmetry breaking becomes an exact Goldstone
boson.

Expanding Eq. \ref{Eq: General LH Potential} to quartic order, the structure of the potential must be of the form
\begin{eqnarray}
\label{Eq: LH Potential Linearized}
V= \lambda_1 (f \phi + [h h])^2 + \lambda_2  (f \phi- [h h])^2 +\cdots
\end{eqnarray}
where  $[h h]$ is a generalized product of Higgs fields, i.e. $h^\dagger h$,  $h^\dagger \tau^a h$, etc.
The first and second terms in the expansion preserve
\begin{eqnarray}
\label{Eq: Shift Symmetries}
\nonumber
\delta_{\epsilon_1} h = \epsilon_1 f &\qquad& \delta_{\epsilon_1} \phi = -[\epsilon_1 h], \\
\delta_{\epsilon_2} h = \epsilon_2 f &\qquad& \delta_{\epsilon_2} \phi = [\epsilon_2 h].
\end{eqnarray}
The primary challenge in constructing a collective quartic is promoting these leading order transformations 
into an algebra that closes.   The field $\phi$ plays a key role in collective symmetry breaking.  When a background
field for $h$ is turned on, $\phi$ acquires a source term and for this reason, we call $\phi$ the ``{\it source field.}''

The transformation properties of the source field are calculated from the product of two Higgs doublets:
$\phi \sim h^\dagger \otimes h, h\otimes h$.
The possible representations of $SU(2)_L\!\times\! U(1)_Y$ for the source field are
\begin{eqnarray}
\phi \sim \mathbf{1}_0, \mathbf{3}_0, \mathbf{1}_{\pm 1}, \mathbf{3}_{\pm 1}.
\end{eqnarray}
When $\phi$ is a singlet, a tadpole can be generated, destroying the stability of the potential.
Additionally, the $\mathbf{1}_{\pm 1}$ requires an antisymmetric product of Higgs fields, therefore one Higgs doublet models of this type of source field are not possible.  
Motivated by the desire for fewer new particles, we consider minimal LH models where minimality is defined as the fewest number of PGB fields.
Additionally we want to have a viable LH potential that does not have quadratic divergences due to tadpoles of singlet fields.
Thus the minimal LH theory  is one that contains a triplet in addition to the 
Higgs doublet -- 7 fields total.  
Some of the  simplest models are
\begin{eqnarray}
\label{Eq: MinimalContent}
\text{dim } G/H =
\begin{cases}
 7  &  \mathbf{2}_\half \oplus \mathbf{3}_0\\
10 & \mathbf{2}_\half \oplus \mathbf{3}_1\\
10 & \mathbf{2}_\half\oplus \mathbf{2}_\half \oplus \mathbf{1}_1\\
10 & \mathbf{2}_{\half,\pm \half_{\text{PQ}}}\oplus \mathbf{1}_{0, 1_{\text{PQ}}}
\end{cases}
\end{eqnarray}
This article shows that none of these models exist without additional fields.
The last example has a global $U(1)_{\text{PQ}}$ that  the Higgs and a neutral singlet are charged under
\begin{eqnarray}
V \simeq \lambda_1 | f \eta + h_1^\dagger h_2|^2 + \lambda_2| f \eta - h_1^\dagger h_2|^2.
\end{eqnarray}
The $SU(6)/Sp(6)$ LH model falls under this category and the global symmetry prevents the singlet from acquiring a dangerous tadpole.
Of course, there could be additional fields that do not participate in the LH potential.  

Sec. \ref{Sec: Criteria} outlines the various mathematical constraints that collective symmetry breaking imposes on the groups $G$, $H$, $\GG_1$ and $\GG_2$.  Sec. \ref{Sec: Models} presents all coset spaces with dimension 14 or less.  After applying the constraints from Sec. \ref{Sec: Criteria}, the smallest possible  models with collective quartics are enumerated.  The smallest model is dimension 12 and is based on the coset $SU(5)/SU(3)\times SU(2)\times U(1)$.

\section{Criteria for Collective Quartics}
\label{Sec: Criteria} 

This section demonstrates the necessary conditions for the Higgs, $h$, and the source field, $\phi$, to have the desired non-linearly realized  symmetries needed to achieve collective symmetry breaking.

The total symmetry of the little Higgs model is $G$ and the linearly realized subgroup is $H$.
In order to not break $G_{\text{EW}}= SU(2)_L\times U(1)_Y$, $G_{\text{EW}}$ must be a subgroup
of $H$.
 The generators of $G$ are normalized to $\Tr T_a T_b = \half \delta_{a b}$.  
 The collective symmetries that protect the Higgs are called $\GG_1$ and $\GG_2$ with their respective generators $T_{\GG_{1,2}}$.
$X_h$ are the generators of the Higgs boson(s) inside of $G/H$ and they transform as $\mathbf{2}_\half$ under $G_{\text{EW}}$.

One of the conditions required for collective symmetry breaking is that all generators of $\GG_{1,2}$ satisfy
\begin{eqnarray}
\label{Eq: Partial Support}
0 \le | \Tr X_h T_{\GG_{1,2}} | < \half,
\end{eqnarray}
where there exists at least one generator that does not trace with $X_h$ to zero.  This condition  is  referred to as ``{\it partial support}" of the Higgs inside the collective symmetry breaking cosets.  
 This condition implies that the generators of the Higgs boson are not completely contained within either $\GG_{1}$ or $\GG_2$
and is essence of partial support.

If the generators of the Higgs boson are completely contained inside of either $\GG_1$ or $\GG_2$, then the generators of the Higgs 
only transforms non-linearly under a single transformation
\begin{eqnarray}
h_i \rightarrow h_i + \epsilon_i f.
\end{eqnarray}
When only one transformation acts on the Higgs generator,
 it is impossible to have two separate operators of the form in Eq. \ref{Eq: LH Potential Linearized} that are guaranteed by symmetries.
 Partial support allows $\GG_1$ and $\GG_2$ to act in distinct ways on the Higgs, creating the possibility of collective symmetry breaking.

Partial support is closely related to the relative embeddings of $H$ into $G$ and of $\GG_1$ and $\GG_2$ into $G$.  This article will demonstrate that partial support is equivalent to the statement that the embedding of $H$ into $G$ is a special embedding relative to the embedding of $\GG_1$ and $\GG_2$ into $G$.    
There are two classes of models that satisfy this criteria.
The first class,  ``{\it special coset},'' is defined as models where $H$ is a special embedding of $G$ and $\GG_1$ and $\GG_2$ are regular embeddings of $G$.
The second class, ``{\it special quartics},'' is defined as models  where
$H$ is a regular embedding of $G$ and $\GG_1$ and $\GG_2$ are special embeddings of $G$.
In some cases, the distinction between these two classifications is blurred, preventing the clean dichotomy of
collective symmetry breaking.  These classifications  are  discussed further in Sec. \ref{Sec: Classification}.

The remaining portion of this section proves this criteria.   Sec. \ref{Sec: Shift Sym} shows how the transformation properties constrain the relation between $H$, $\GG_1$ and $\GG_2$ and results in the requirement of ``partial support.''  Sec. \ref{Sec: Potential} shows how a class of quartic couplings can is related to the structures presented in Sec. \ref{Sec: Shift Sym}.  While the example presented in Sec. \ref{Sec: Potential} is the simplest example of a collective quartic, it appears in several models, including the Littlest Higgs and the model introduced in this article.  Finally, Sec. \ref{Sec: Classification} relates partial support to special embeddings of subgroups.

\subsection{Source Fields and Shift Symmetries}
\label{Sec: Shift Sym}

In Little Higgs models, the Higgs is a pseudo-Goldstone boson created by breaking the group $G$ down to the subgroup $H$ via a vev, $\Sigma$.    The Goldstones non-linearly realize a symmetry.  In order for the Higgs to acquire a potential of the form in Eq. \ref{Eq: General LH Potential}, it is first necessary to classify how subgroups of $G$ act upon $\Sigma$.   Little Higgs models are restricted to the case where the Higgs and source field transform nonlinearly under two distinct groups, $\GG_1$ and $\GG_2$ as shown in Eq. \ref{Eq: Shift Symmetries}.

The generators of  $G/H$ are broken generators and labeled by $X$.  The generators of $H$ are unbroken generators, $T_H$.   Unbroken generators act on the vev of $\Sigma$ and vanish
\begin{eqnarray}
T_H \langle\Sigma\rangle = 0 .
\end{eqnarray}
The unbroken generators of $G$ will play an important role in elucidating the role of collective symmetry breaking.

In LH theories, the Higgs, $h$, has a shift symmetry under $\GG_1$ and $\GG_2$ while the source field, $\phi$, transforms proportionally to the Higgs (see Eq. \ref{Eq: Shift Symmetries}).  These transformations imply constraints on $\GG_1$ and $\GG_2$.  Parameterizing the broken directions as $\pi$ and performing a transformation under $\GG_1$ gives 
\begin{eqnarray}
e^{i\epsilon T_{\GG_1}} e^{i\pi/f} \langle\Sigma\rangle =   e^{i\pi'/f} \langle \Sigma\rangle
\end{eqnarray}
 where $T_{\GG_1}$ are the generators of $\GG_1$, and the equation is suitably generalized for higher tensor representations of $\Sigma$.  
 The Baker-Campbell-Hausdorff formula gives 
 \begin{eqnarray}
\label{Eq: BCH}
\pi' = \pi + \epsilon f T_{\GG_1} +\frac{i}{2}[\epsilon T_{\GG_1},\pi] + \OO(\epsilon^2,\pi^2).  
 \end{eqnarray}
 The Higgs shift arises from the first term in the expansion.  
 
\begin{figure}[tb] 
   \centering
   \includegraphics[width=4in]{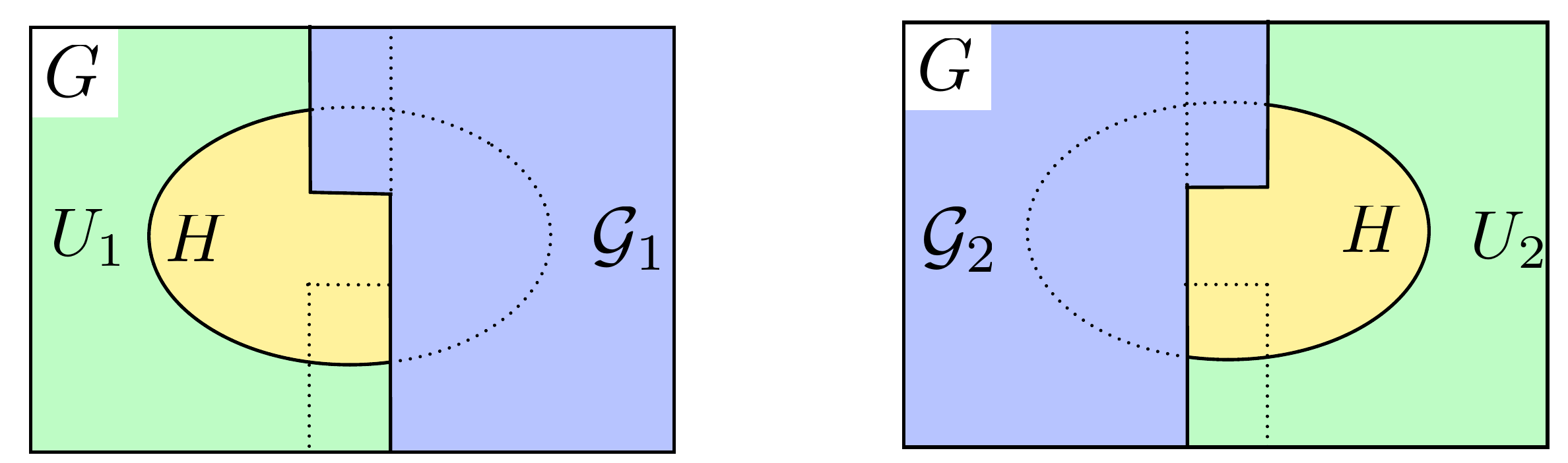} 
  \caption{A schematic diagram of the different decompositions of the generators of $G$.  The left shows the decomposition of $G$ under $\GG_1$ while the right shows the decomposition under $\GG_2$.  The Higgs is in an admixture of the generators of $H$ and the generators of $\GG_1$ and $\GG_2$ while the source field, $\phi$, falls outside of $\GG_1$, $\GG_2$ and $H$.}
  \label{Fig: Venn}
\end{figure}

\begin{figure}[tb] 
   \centering
   \includegraphics[width=4in]{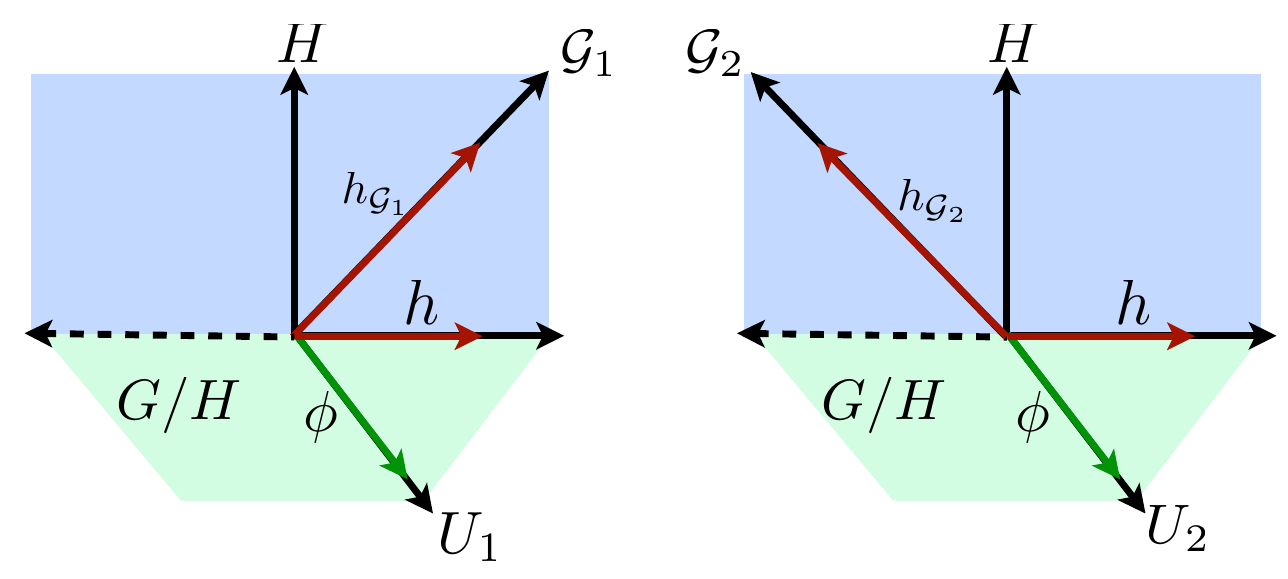} 
   \caption{Another schematic diagram of the multidimensional space spanned by the generators of $G$.  The Higgs is uniquely decomposed as a sum of the non-orthogonal generators of $\GG_1$, $H$ and $U_1$.  The generators of $U_1$ are those for which the $G/H$ and $G/\GG_1$ axes coincide.  Instead of the projection onto the usual orthonormal basis, the Higgs is projected onto the nonorthogonal spaces of $\GG_1$ and $H$.}
   \label{Fig: Linear}
\end{figure}

The relation of $\pi'$  to $\pi$ is determined by expanding the generators of $G$ into three terms:  $T_{\GG_1}$, the generators of $\GG_1$; $T_H$, the generators of $H$; and $X_{U_1}$,  the remaining generators of $G$, where
\begin{eqnarray}
U_1 =G/H \cap G/\GG_1.
\end{eqnarray}  
The generators of $\GG_1$ provide the shift symmetry for the Higgs boson.  The generators of $\GG_1$ are an admixture of generators in $H$ and $G/H$.
Fig. \ref{Fig: Venn} and  Fig. \ref{Fig: Linear} show this decomposition diagrammatically. 

 The generators of the Higgs are decomposed as
\begin{eqnarray}
\label{Eq: Higgs Basis}
X_h = c^h_{\GG_1} T^h_{\GG_1} + c^h_H T^h_H + c^h_{U_1} X^h_{U_1}
\end{eqnarray}
where the $c$s are constants with implied indices.  The $\OO(\epsilon)$ transformation of the Higgs in the $T^h_{\GG_1}$ direction is
\begin{eqnarray}
 h' X_h = h X_h + \epsilon f T^h_{\GG_1}
 =
  (h c^h_{\GG_1} + \epsilon f) T^h_{\GG_1}  + hc^h_{U_1} X^h_{U_1} + h c^h_H T^h_H .
\end{eqnarray}
At linear order, the last term vanishes when acting upon $\langle \Sigma \rangle$.
This transformation is interpreted as the Higgs obtaining a shift symmetry only if $c^h_{\GG_1}\ne 0$ and $c^h_{U_1} = 0$.

The source field, $\phi$, should not acquire a shift symmetry under $\GG_1$. Therefore, when decomposing $\phi$ in 
an analogous manner to the Higgs in Eq. \ref{Eq: Higgs Basis},
\begin{eqnarray}
X_\phi = c^\phi_{\GG_1} T^\phi_{\GG_1} + c^\phi_H T^\phi_H + c^\phi_{U_1} X^\phi_{U_1}.
\end{eqnarray}
If $c^\phi_{\GG_1} \ne 0$, then it is possible to do a transformation in that direction; however, this symmetry transformation would prevent $\phi$ from acting as the source field in \mbox{Eq. \ref{Eq: LH Potential Linearized}}.
Therefore, $c^\phi_{\GG_1} = 0$ and since $\phi$ cannot only live inside of $H$,  $c^\phi_{U_1} \ne 0$.  
Since, $U_1$ is orthogonal to $H$ (see Fig. \ref{Fig: Linear})  $c^\phi_H$ vanishes.

The desired transformation of $\phi$ is of the form 
 \begin{eqnarray}
 \delta_\epsilon \phi \propto [\epsilon h].
\end{eqnarray}
The second order term in Eq. \ref{Eq: BCH} constrains the transformation properties of $\phi$ up to unbroken generators
 \begin{eqnarray}
\label{Eq: Phi Decomposition}
 X_{\phi} \propto  [T^h_{\GG_1},X_h] \propto [T^h_{\GG_1} , c_{\GG_1}^h T^h_{\GG_1}] + [T^h_{\GG_1}, c_H^h T^h_H ] 
               \propto X^\phi_{U_1}
\end{eqnarray}
The first term closes onto an element of $\GG_1$, which is orthogonal to $\phi$.  Thus the only way that Eq. \ref{Eq: Phi Decomposition} can be nonzero is if $c_H^h \ne 0$.  The only consistent transformation is
\begin{eqnarray}
\nonumber
[T^h_{\GG_1}, c_H^h T^h_H ] &=& c_H ^h  d^{U_1}_{\GG_1\, H} X_{U_1} + c_H^h d^H_{\GG_1\, H} T_H + c^h_H d^{\GG_1}_{\GG_1 H} T_{\GG_1}\\
 \,[T^h_{\GG_1}, c_{\GG_1}^h T^h_{\GG_1} ] &=&  c_{\GG_1}^h d^{\GG_1}_{\GG_1 \,\GG_1} T_{\GG_1} .
\end{eqnarray}
In order for this to be proportional to $X_{U_1}$ 
\begin{eqnarray}
c_{\GG_1}^h d^{\GG_1}_{\GG_1 \,\GG_1} T_{\GG_1} + c^h_H d^{\GG_1}_{\GG_1 H}  T_{\GG_1}=0 
\end{eqnarray}
so that $c_{\GG_1}^h d^{\GG_1}_{\GG_1 \,\GG_1} = - c^h_H d^{\GG_1}_{\GG_1 H}$.  In principle, $d^{\GG_1}_{\GG_1 \,\GG_1}, d^{\GG_1}_{\GG_1 H} \ne 0$; however there are no known examples of simple group LH model with this property.

This line of argument has shown that in order for PGBs from a coset to form a potential of the form
\begin{eqnarray}
V_1 = \lambda_1( \phi f + [hh])^2 + \cdots,
\end{eqnarray}
where the form of this potential is guaranteed by a symmetry, $\GG_1$, it is necessary that the generators of $\GG_1$ be an admixture of generators in $H$ and $G/H$.  Specifically, the generators in the direction of the Higgs boson, $X_h$, are a linear combination of those in $\GG_1$ and $H$.
 This demonstrates that the normalized generators, $X_h$, must trace to less than $\half$ against all generators of $\GG_1$, proving the requirement of partial support given in Eq. \ref{Eq: Partial Support}. 
A similar set of arguments applies for $\GG_2$.

Two Higgs doublet models have slightly different source field transformations:
 \begin{eqnarray}
 \delta \phi \propto [\epsilon_1 h_2] +  [\epsilon_2 h_1].
\end{eqnarray}
The corresponding conditions on the source field are obtained by requiring that the two terms in the commutator come from two different Higgs.  The relevant terms are 
 \begin{eqnarray}
 X_{\phi} \propto  [T^{h_1}_{\GG_1}, c_H^{h_2} T^{h_2}_H ] + [T^{h_2}_{\GG_1}, c_H^{h_1} T^{h_1}_H ] .
\end{eqnarray}
Other than additional complexity, there are no new features to consider for this class of models.

\subsection{Constructing a Potential}
\label{Sec: Potential}

This section constructs the collective symmetry breaking potential with minimal  field content and  is  representative of more general models.  The minimal model involves groups that have two sets of generators in $G$ that transform as doublets.
One of these doublets is an element of $H$ and the other is an element of $G/H$ and  is  the physical pseudo-Goldstone Higgs boson.

Using the results from the previous section, it is necessary for  $\GG_1$ and $\GG_2$ to act in distinct ways, {\em i.e.} Eq. \ref{Eq: Shift Symmetries}.
Thus, the transformation of $\phi$ under $\GG_1$ and $\GG_2$ must be related
\begin{eqnarray}
\label{Eq: Phi Transformation}
\delta\phi X^\phi_{U_1} \propto c^h_{\GG_1} [T^h_{\GG_1},X_{h}] = - c^h_{\GG_2} [T^h_{\GG_2},X_{h}] .
\end{eqnarray}
Having constructed the symmetry pattern necessary for collective symmetry breaking, the next step is to construct
the potential.  

As an example, the collective symmetry breaking potential that is realized in some of the most common Little Higgs models,
such as the Littlest Higgs, can be constructed using the tools above.  
This structure is present in the model introduced in Sec. \ref{Sec: New Model}.  Suppose $\GG_1$ has 
\begin{eqnarray}
T_{1} \in \GG_1 \qquad T_{2} \in G/\GG_1,
\end{eqnarray}
Where $T_1$ and $T_2$ are doublet generators.  $X_{h}$ must be a combination of $T_{1}$ and $T_{2}$ to get partial support.
\begin{eqnarray}  
X_{h}=\frac{1}{\sqrt{2}}(T_{1}+ T_{2})  \in G/H \qquad
T^h_{H}=\frac{1}{\sqrt{2}}(T_{1}- T_{2}) \in H.
\end{eqnarray}
Under a $\GG_1$ transformation, the Higgs transforms nonlinearly 
\begin{eqnarray}
h' X_{h}  \propto   h X_{h} + \epsilon T_{1} 
 \propto   ( h + \frac{\epsilon}{\sqrt{2}} ) X_{h} +  \frac{\epsilon}{\sqrt{2}} T^h_{H}.
\end{eqnarray}

The Baker-Campbell-Hausdorff formula shows that the source field, $\phi$, appears in the commutator of the generators of $\GG_1$ with the Higgs 
\begin{eqnarray}
X_{\phi} \propto [T_{1},X_{h}] \in G/H.  
\end{eqnarray}
$X_{\phi}$ must lie outside of $\GG_1$ to avoid a shift symmetry:
\begin{eqnarray}
X_{\phi} \propto [T_{1},X_{h}] \propto [T_{1},T_{1}] + [T_{1},T_{2}].
\end{eqnarray}
Because the commutator of two generators of $\GG_1$ is either 0 or lies within the root space of $\GG_1$, this requirement leads to the condition
\begin{eqnarray}
[T_{1},T_{1}] = \alpha_1 T_H .
\end{eqnarray}
 ($T_1$ is a doublet, a set of 4 generators, so its commutator does not necessarily vanish).  Thus $X_{\phi}$ acquires the desired shift symmetry shown in Eq. \ref{Eq: Shift Symmetries}.

These shift symmetries show that there is a $\GG_1$ covariant operator, $\OO_1$, that when expanded becomes
\begin{eqnarray}
\label{Eq: Half Potential}
\OO_1 = \PP_1 \Sigma \PP_1 = \phi/f + [hh]/f^2  + \cdots,
\end{eqnarray}
where $\PP_1$ is a projection operator that preserves $\GG_1$.
Thus, the desired potential is
\begin{eqnarray}
V_1 = \lambda_1 f^4 \Tr \OO_1 \OO_1^\dagger .
\end{eqnarray}

The next step is to consider the source of the minus sign difference between the two quartics in this example.  The correction to $\phi$ is proportional to $[T_{1},T_{2}]$.  This correction is written more suggestively as
\begin{eqnarray}
[T_{\GG_1},T_{G/\GG_1}].
\end{eqnarray}
The only construction that gives a potential with the opposite sign in the quartic of Eq. \ref{Eq: Half Potential} requires flipping the generators that are in $G/\GG_1$ with the generators in $\GG_1$:
\begin{eqnarray}
T_{2} \in \GG_2 \qquad T_{1} \in G/\GG_2 
\qquad
[T_{2},T_{2}] = \alpha_2 T_H.
\end{eqnarray}
As before, the Higgs acquires a shift symmetry.  
The correction to the quartic is $[T_{2},T_{1}]$ so there is a $\GG_2$ covariant operator,  $\OO_2$,
\begin{eqnarray}
\label{Eq: Half Potential}
\OO_2 = \PP_2 \Sigma \PP_2 = \phi /f- [hh]/f^2  + \cdots,
\end{eqnarray}
where $\PP_2$ is a projection operator that preserves $\GG_2$.
  
This example shows one straightforward, but not unique, way to fulfill the otherwise obscure Eq. \ref{Eq: Phi Transformation}.  If the Higgs can be expressed as
\begin{eqnarray}
X_h = c_1 T_{\GG_1 \cap G/\GG_2} + c_2 T_{\GG_2 \cap G/\GG_1},
\end{eqnarray}
then the opposite-sign shift symmetries are automatically satisfied.  
The minimal models $SU(5)/SO(5)$, $SU(6)/Sp(6)$ and $SU(5)/SU(3)\times SU(2)\times U(1)$ all use this mechanism to obtain the difference in sign between the two quartic terms.

\subsection{Relation of Partial Support to Special Embeddings}
\label{Sec: Classification}

Special embeddings of Lie groups are subgroups whose roots are not the roots of the full group.
More colloquially, special embeddings are those subgroups whose Dynkin diagrams are not created by removing nodes of the extended Dynkin diagrams. A familiar class of special embeddings are $SO(2n-1)$ in $SO(2n)$.  
``Partial support'' implies that either $H$ or both $\GG_{1}$ and $\GG_2$ must be special embeddings of $G$.   
The broken generators of the Higgs must lie both inside and outside of the $\GG$s to get partial support.  In the basis chosen by the root space of $G$, the roots of a regular embedding are simple subsets of the roots of $G$ up to a mixing of the Cartan subalgebra. 

The need for a special embedding is seen directly from Fig. \ref{Fig: Linear}.  
The condition of partial support in Eq. \ref{Eq: Higgs Basis} is 
\begin{eqnarray}
c^h_{\GG_1} \ne 0 \quad \text{ and } \quad c^h_H \ne 0. 
\end{eqnarray}
 This decomposes an element in $X_{G/H}$ into $T_H$ and $T_{\GG_1}$.  $X_{G/H}$ and $T_H$ are orthogonal so the only way that the decomposition can have a nonzero projection onto $T_H$ is if the generators of $T_{\GG_1}$ and $T_H$ are not orthogonal; in other words, the two are relatively special embeddings.  This argument is seen pictorially from Fig. \ref{Fig: Linear}; the only way to have a nonzero projection of the Higgs onto $T_H$ is to have the axis of $\GG_1$ be at an angle with respect to the axis of $H$.  $\GG_1$ and $H$ are relatively special embeddings.  As a result, one of the two must be a special embedding in $G$.

Collective quartics require specifying the embeddings of three groups ($H$, $\GG_1$ and $\GG_2$) into $G$.
It is possible to have two apparently regular embeddings where  the choice of basis is not mutually compatible. 
This mutual incompatibility results in the roots of the second group being a linear combination of the roots of the generators of the first group.    For instance, consider two $SU(2)$ subgroups of $SU(3)$.  The first subgroup is always chosen to be transformations of the form
\begin{eqnarray}
U_{SU(2)_1} \sim \left(\begin{array}{ccc} \square & \square & 0\\\square & \square & 0\\0&0&1\end{array}\right);
\end{eqnarray}
however, the second $SU(N)$ subgroup is of the form
 \begin{eqnarray}
 \label{Eq: Relative Special}
U_{SU(2)_2} \sim \left(\begin{array}{ccc} 1&0&0\\0&c_\theta& s_\theta\\0 & -s_\theta & c_\theta\end{array}\right)
 U_{SU(2)_1} 
\left(\begin{array}{ccc} 1&0&0\\0&c_\theta& -s_\theta\\0 & s_\theta & c_\theta\end{array}\right),
\end{eqnarray}
where $\theta$ is a fixed value defining the relative embedding of $SU(2)_1$ to $SU(2)_2$.  
If $\sin 2\theta \ne 0$, then $SU(2)_2$ is a relatively special embedding to $SU(2)_1$. 
This simple example shows that while $H$ and $\GG$ must be relatively special embeddings, it is not always
clear which one is a special embedding in $G$.

The requirement that a collective quartic requires special embeddings is the strongest constraint in building minimal LH models.
Large coset spaces admit  many structures of the form described in Eq. \ref{Eq: Relative Special}  and therefore the Higgs can be spread throughout the coset.   The ability to support the Higgs in multiple location means that large cosets usually admit relatively special embeddings and the condition of partial support does not constrain larger models.
 Small dimensional cosets are much more constrained and spreading the Higgs out over multiple generators restricts the possible candidate theories.   
 Requiring the $\GG$s to act linearly on the electroweak generators of the Standard Models further restricts possible models.

\section{Models}
\label{Sec: Models}

This section categorizes Little Higgs models that derive from a simple group that have collective quartic couplings with the fewest number of pseudo-Goldstone bosons.  
The smallest Little Higgs models known are $SU(5)/SO(5)$ and $SU(6)/Sp(6)$ and each have 14 Goldstone bosons.  This section
only considers theories with no more than 14 Goldstones.  Additionally, $H$ must have rank greater than or equal to 2
and $G/H$ needs to be at least dimension 7.  The list below contains all cosets that satisfy these constraints:
\begin{itemize}
\item Dim 7:  $a_3/a_2$, $b_3/g_2$, $d_4/b_3$
\item  Dim 8: $g_2/d_2$, $c_3/c_2\!\times\! c_1$, $a_3/a_1\!\times\! a_1 \!\times\! a_0$, 
$b_4/d_4$, $a_4/a_3\!\times\! a_0$
\item Dim 9: $a_3/d_2$, $a_4/a_3$, $d_5/b_4$
\item Dim 10: $b_3/b_2\!\times\! a_0$, $b_5/d_5$, $a_5/a_4\!\times\! a_0$
\item Dim 11: $b_3/b_2$, $c_3/c_2$, $a_5/a_4$, $d_6/b_5$
\item Dim 12:
$c_3/a_2\!\times\! a_0 $, $c_4/c_3\!\times\! c_1 $, $b_6/d_6$,
$b_3/d_2\!\times\! b_1 $, $d_4/a_3\!\times\! a_0 $, $a_6/a_5\!\times\! a_0$,
$c_3/c_1 \!\times\! c_1\!\times\! c_1 $, $a_4/a_2\!\times\! a_1 \!\times\! a_0$, $b_3/a_2 \!\times\! a_0$
\item Dim 13:
 $c_3/a_2 $, $d_4/a_3$, $a_6/a_5$, $d_7/b_6$, 
$b_3/a_2$, $a_4/a_2 \!\times\! a_1$
\item Dim 14: $a_4/b_2$, $a_5/c_3$, $b_4/b_3\!\times\! a_0$, $b_7/d_7$,
$d_4/g_2$, $a_7/a_6\!\times\! a_0$
\end{itemize}
where the groups are labeled by their standard Dynkin name (\eg $a_3 = SU(4)$) and $a_0 = U(1)$.
The next step in the classification is to find special subgroups of $G$.   If $H$ is special, then these theories are called ``{\it special cosets}.''
Otherwise, it is necessary to find special subgroups for $\GG_1$ and $\GG_2$, denoted as ``{\it special quartics}.'' 

\subsection{Special Cosets}

The first class of models where there is a collective quartic are those in which $H$ is a special embedding
of $G$, but $\GG_1$ and $\GG_2$ are regular embeddings
of $G$.  From the list presented above, the special embeddings are
\begin{eqnarray}
G/H  = \begin{cases}
SO(n)/G_2,& n=7,8\\
SU(n)/SO(n), & n=4,5 \\
SU(2n)/Sp(2n),& n=3\\
SO(2n)/SO(2n-1) & n=4, \ldots, 7
\end{cases} .
\end{eqnarray}

\begin{itemize}
\item
$SO(7)/G_2$ is simple.  The root space of $SO(7)$ only contains a single doublet.  Since at least 2 doublets are needed for partial support, this model is ruled out.  For the case $SO(8)/G_2$, $SU(2)_L \times U(1)_Y \subset SU(2)_L \times SU(2)_R \subset SO(4) \times SO(3)$.  The $SU(2)_L$ is a diagonal combination of one of the $SU(2)$s in the $SO(4)$ and the $SO(3)$.  $SO(8)$ contains two doublets but they do not commute into the triplet, which rules out this model.
\item
$SU(4)/SO(4)$'s field content only admits Higgs doublets and neutral singlets and therefore has dangerous singlets.
\item
$SO(2n)/SO(2n-1)$ (including $SU(4)/Sp(4) \simeq SO(6)/SO(5)$ ) suffer from the dangerous singlet problem \cite{Schmaltz:2008vd}.  In general, there exist multiple doublets.  Simple computation shows that these doublets commute into a singlet.  Giving the singlet a charge would also give one of the doublets a different charge, theerby preventing the mixing needed for a LH model.
\end{itemize}

The models with the fewest number of fields that have collective quartics are $SU(5)/SO(5)$ and $SU(6)/Sp(6)$ \cite{ArkaniHamed:2002qy,Low:2002ws}.  
These are well known examples of Little Higgs models.  

\subsection{Special Quartics}

The second type of collective quartic arises when $\GG_1$ and $\GG_2$ are special embeddings of $G$.   These cases are easy to identify by checking all possible special embeddings that might contain the SM.  In almost every case, the special embeddings do not contain the SM.  Notable exceptions are $SU(N)/SO(N)$ where the $SO(N)$ contains the diagonal $SU(2) \times SU(2)$ of the $SU(N)$.  For $SU(2N)/Sp(2N)$, the $Sp(2N)$ also contains a diagonal subgroup of $SU(2N)$.

One coset that admits a special quartic is $SU(5)/SU(3)\times SU(2)\times U(1)$ where there are two overlapping $SO(5)$s that generate the Higgs mass.
This model is discussed in some depth below.  This model illustrates a duality between special embedding and special quartics.  In the Littlest Higgs, the unbroken group was $H=SO(5)$, which is a special embedding, and the non-linearly realized group generating the Higgs quartic is $SU(3)\times SU(2)\times U(1)$, which is a regular embedding.  A special quartic is generated by interchanging the non-linearly realized groups with the unbroken global symmetry group.
There must be two distinct $SO(5)$ embeddings in order for this duality to hold, but frequently there is a parity that guarantees this is the case.

The more challenging set of special quartics to identify are those whose $\GG_1$ and $\GG_2$ are either regular or special embeddings depending on the relative alignment of the $\GG$s to $H$.  If there are multiple ways of embedding $\GG_1$ in $G$, then linear combinations of these embeddings can also satisfy the algebra of $\GG_1$.
An example is the Simple Custodially Symmetric Little Higgs ($SO(9)/SO(5)\times SO(4)$) \cite{Skiba:2003yf}.  The collective quartic arises from two $\GG_{1,2} = SO(5)\times SO(4)$.  On the surface, this model looks like a case where both $\GG_{1,2}$ and $H$ are regular embeddings. Using the standard basis of roots for $SO(9)$ \cite{Georgi:1982jb}, the roots of $H$ are sums of the roots of $G$ and $H$ is a special embedding.  Alternatively, it is possible to choose a basis where $H$ is regular, however $\GG_{1,2}$ are then special.
 $SO(9)/SO(5)\times SO(4)$ shows that the distinction between special cosets and special quartics is not always well-defined.

 \subsubsection{$SU(5)/SU(3)\times SU(2)\times U(1)$}
 \label{Sec: New Model}
 
The smallest viable special quartics model is $SU(5)/SU(3)\times SU(2)\times U(1)$ and  it is the ``dual'' of the Littlest Higgs where the special embedding of the unbroken symmetry is interchanged with the regular embedding of the non-linearly realized groups.  Much of the structure between the two theories is the same. 
The generators of the SM are 
\begin{eqnarray}
\tau^a = \frac{1}{2 \sqrt{2}}
\begin{pmatrix}
\sigma^a & 0 & 0 \\
0 & 0 & 0 \\
0 & 0 & -\sigma^{a *}
\end{pmatrix} \qquad
Y = \frac{1}{2}
\begin{pmatrix}
\identity_2 & 0 & 0 \\
0 & 0 & 0 \\
0 & 0 & -\identity_2
\end{pmatrix}.
\end{eqnarray}
The breaking is done by an adjoint field $\Sigma$, where
\begin{eqnarray}
\langle\Sigma\rangle = \Sigma_0 = \frac{1}{2 \sqrt{15}}
\begin{pmatrix}
2\, \identity_2 & 0 & 0 \\
0 & 2& 0 \\
0 & 0 & -3\,\identity_2
\end{pmatrix} .
\end{eqnarray}
These broken directions are parameterized as 
\begin{eqnarray}
\Sigma = e^{i \pi/f}\; \Sigma_0\; e^{-i \pi/f},
\end{eqnarray}
where the coset space contains a Higgs, a charged triplet as the source field and a charged scalar and  
is decomposes as
\begin{eqnarray}
\label{Eq: 5/321Basis}
\pi = 
\begin{pmatrix}
0 & 0 & \phi^{lm} + \epsilon^{lm} s \\
0 & 0 & h^m \\
\phi_{i j}^\dagger + \epsilon_{ij} s^\dagger & h_i^\dagger & 0
\end{pmatrix}.
\end{eqnarray}
The fields transform as
\begin{eqnarray}
h \sim \mathbf{2}_\half \qquad \phi \sim \mathbf{3}_1 \qquad s \sim \mathbf{1}_1
\end{eqnarray}
under $SU(2)_L\times U(1)_Y$.

The leading order Lagrangian is
\begin{eqnarray}
\label{Eq: Kin}
\LL_{\text{kin}} = \frac{6 f^2}{5} \Tr |D_\mu \Sigma|^2,
\end{eqnarray}
where the constant $\frac{6}{5}$ is chosen for canonical kinetic terms for the definition in Eq. \ref{Eq: 5/321Basis}.

Within the unbroken subgroup, $H$, there exists a doublet that commutes with the Higgs to give the triplet and scalar.  This doublet is
\begin{eqnarray}
\tilde{h} \;T_{H}^h=\begin{pmatrix}
0 & \tilde{h} & 0 \\
\tilde{h}^\dagger & 0 & 0 \\
0 & 0 & 0
\end{pmatrix}
\end{eqnarray}
and participates in collective symmetry breaking together with the Higgs; this doublet is the $T^h_H$ that played an important role in the previous section.  While there is no regular embedding that contains a linear combination of the two Higgses, there is a special embedding that does, $SO(5)$.  There exist two different $SO(5)$s that contain the Higgs.  

The two $SO(5)$s that contain the Higgs are constructed from the generators that obey
\begin{eqnarray}
T_{1 a} P_1 + P_1 T_{1 a}^T = 0 \qquad
T_{2 a} P_2 + P_2 T_{2 a}^T = 0, 
\end{eqnarray}
where
\begin{eqnarray}
P_1 = \begin{pmatrix}
0  & 0  & \identity_2\\
0 & 1 & 0\\
\identity_2 & 0 & 0
\end{pmatrix} 
\qquad
P_2 = \begin{pmatrix}
0  & 0  & \identity_2\\
0 & -1 & 0\\
\identity_2 & 0 & 0
\end{pmatrix}.
\end{eqnarray}
The generators that satisfy these equations are exactly the same, except for the four generators 
\begin{eqnarray}
T_{1,2}= \frac{1}{\sqrt{2}} ( X_{G/H}^h \pm T^h_H ) = \frac{1}{\sqrt{2}}
\begin{pmatrix}
0 & \pm h^T & 0 \\
\pm h^* & 0 & h \\
0 & h^\dagger & 0
\end{pmatrix}
\end{eqnarray}
with the $\pm$ depending on which $P_i$ is used.  These generators are $T_1$ and $T_2$ used in the previous example.   The scalar is contained in both of the $SO(5)$s and so receives a Higgs-like shift symmetry.  Under action of $T_{1}$, there are two relevant sets of transformations: those that, to leading order, shift the Higgs boson denoted by $\epsilon_1$, and those that shift the charged singlet denoted by $\tilde{\epsilon}_1$.  These act in the following manner:
\begin{eqnarray}
\delta_{\epsilon_1} 
\begin{cases} h^i \\\phi^{ij} \\s\end{cases}
=
\begin{cases}
\epsilon_{1}^i f+\cdots \\
 - \frac{i}{4} (\epsilon_1^i h^j + \epsilon_1^j h^i) + \cdots\\
0\\
\end{cases} \qquad
\delta_{\tilde{\epsilon}_1} 
\begin{cases} h^i \\\phi^{ij} \\s\end{cases}
=
\begin{cases}
0 \\
0\\
\tilde{\epsilon}_1 f  + \cdots \\
\end{cases} 
\end{eqnarray}
and under the action of the two transformations in $T_{2 a}$, the fields transform as
\begin{eqnarray}
\delta_{\epsilon_2} 
\begin{cases} h^i \\\phi^{ij} \\s\end{cases}
=
\begin{cases}
\epsilon_{2}^i f+\cdots \\
\frac{i}{4} (\epsilon_2^i h^j + \epsilon_2^j h^i) + \cdots\\
0\\
\end{cases} \qquad
\delta_{\tilde{\epsilon}_2} 
\begin{cases} h^i \\\phi^{ij} \\s\end{cases}
=
\begin{cases}
0\\
0\\
\tilde{\epsilon}_2 f  + \cdots\\
\end{cases} 
\end{eqnarray}
These transformations allow for the expected Little Higgs form of the potential:
\begin{eqnarray}
\label{Eq: 5/321Potential}
\nonumber
V &=& \frac{3}{5} \lambda_1 f^4 \Tr P_1 \Sigma P_1 \Sigma^*
   + \frac{3}{5} \lambda_2 f^4 \Tr P_2 \Sigma P_2 \Sigma^* \\
   &=&  \lambda_1 (f \phi_{i j} + \frac{i}{2} h_i h_j )^2 
   + \lambda_2  (f \phi_{i j} - \frac{i}{2} h_i h_j  )^2 + \cdots
\end{eqnarray}
Upon integrating out the triplet, an independent Higgs quartic is formed.  The gauge quadratic divergences are not canceled, so this model is an Intermediate Higgs model but with a quartic coupling and no dangerous singlets.

The Higgs quartic is 
\begin{eqnarray}
\label{Eq: Higgs Quartic}
\lambda^{-1}= \lambda_1^{-1}+ \lambda_2^{-1}
\end{eqnarray}  
and the mass of the triplet is 
\begin{eqnarray}
\label{Eq: TripletMass}
m_{\phi}^2= (\lambda_1+\lambda_2)f^2  = 4 \lambda f^2/\sin^2 2\theta_\lambda,
\end{eqnarray} 
where $\tan^2 \theta_\lambda = \lambda_1/\lambda_2$.
This theory does not have custodial $SU(2)$ and there are two contributions to the $T$ parameter.  The first arises from the triplet vev and the second arises from the expansion of the kinetic term of Eq. \ref{Eq: Kin}.  After integrating out  $\phi$, the operator that contributes to the $T$ parameter is induced:
\begin{eqnarray}
\LL_{\text{eff}} = \frac{1}{\Lambda_T^2} |h^\dagger D_\mu h|^2,
\end{eqnarray}
where $\Lambda_T$ is
\begin{eqnarray}
\frac{1}{\Lambda_T^2}=
\frac{-1}{f^2} \left( 1 - \frac{1}{4} \frac{(\lambda_1 -\lambda_2)^2}{(\lambda_1+\lambda_2)^2} \right)  
= \frac{-1}{f^2}\left(
1 -\frac{1}{4} \cos^22\theta_\lambda
\right) .
\end{eqnarray}
Since the bounds the $T$ parameter require $\Lambda_T \gsim 3 \TeV$, these contributions  set a bound on $f \gsim 3 \TeV$ unless there is another contribution to the $T$ parameter.  The effects from the $T$ parameter can be lessened by having a moderately large Higgs mass which contributes negatively to the $T$ parameter and positively to the $S$ parameter.   The degree to which this can be employed depends upon other contributions to the $S$ parameter.
   
Other precision electroweak observables are model-dependent and require specifying  how the top quarks cancel the quadratic divergences, how the operators in Eq. \ref{Eq: 5/321Potential} are generated and how the gauge quadratic divergences are cut off.  
Since there are no extra gauge bosons, there are no extra mixings that induce the $S$ parameter.  This theory might be embedded into a more complete moose diagram or an AdS construction  \cite{Piai:2004yb,Thaler:2005en,Thaler:2005kr,Cheng:2006ht,Gregoire:2002ra,Csaki:2008se,Contino:2006nn,Contino:2003ve,Agashe:2004rs,Katz:2003sn}.  The masses of the new vector bosons are now independent from the masses of the top partners and can have masses greater than 3 TeV (required by the $S$ parameter) without forcing the top partners to become heavy or pushing the limits of perturbativity.

\section{Conclusion}

In Little Higgs models, the Higgs is a pseudo-Goldstone boson that results from the breaking of the group $G$ to $H$.  Together with the idea of collective symmetry breaking, Little Higgs models generate a quartic term for the Higgs without generating a mass term, thereby avoiding problems with fine tuning and overly light Higgs.  The two symmetries that participate in the collective symmetry breaking are $\GG_1$ and $\GG_2$.    The key requirement in constructing minimal Little Higgs models is partial support: the Higgs must be acted upon by both $\GG_1$ and $\GG_2$.  Partial support is equivalent to the embedding of $H$ into $G$ being relatively special to the embedding of $\GG_1$ and $\GG_2$ into $G$.  

The requirement of partial support leads to the classification of the smallest possible LH models by exhaustion.  Specifically, none of the minimal field content models listed in Eq. \ref{Eq: MinimalContent} existed.  The smallest possible LH model  with a collective quartic is $SU(5)/SU(3)\times SU(2)\times U(1)$ using two different $SO(5)$s as $\GG_1$ and $\GG_2$.  Because of the absence of particles to act as gauge boson longitudinal modes for an enlarged gauge group, this LH model must be an Intermediate Higgs model \cite{Katz:2005au}.  The source field for this model has electroweak quantum numbers of $\mathbf{3}_1$.  In addition to the source field, $SU(5)/SU(3)\times SU(2)\times U(1)$ also contains a charged scalar that does not participate in the collective symmetry breaking.  The absence of custodial $SU(2)$ is a drawback to this model, however, fully specifying a TeV-scale model (including the top partners) could aid in reducing the $T$ parameter.
In fact, many models with top partners can have large positive contributions to the $T$ parameter (see for instance \cite{Gregoire:2003kr,Baumgart:2007js,Barbieri:2007bh}) and this negative contribution might be desirable.  Further problems with electroweak precision tests, most importantly the $S$ parameter, are alleviated by the minimal structure of the model \cite{Csaki:2003si,Kribs:2003yu,Kilic:2003mq}.  The relevant measurements for collider physics depend upon the origin of the top quark's Yukawa coupling.  While the origin of the Yukawa coupling is not explored here, many of the features are similar to other LH models and can be measured at colliders \cite{Perelstein:2003wd,Han:2003wu,Perelstein:2005ka,Mrazek:2009yu,Asakawa:2009qb,Han:2008xb,Kong:2007uu,Barcelo:2007if}.

The classification presented in this article can be used to investigate the higher order interactions of the Higgs boson arising from the non-linear sigma model structure \cite{Low:2009di,Dobado:2009sc}. 
The nature of radiatively driven electroweak symmetry breaking in Little Higgs models may be further explored using the general methods in this paper  \cite{Coleman:1973jx,Kaplan:1983fs,Banks:1984gj,ArkaniHamed:2001nc,Grinstein:2008kt}.  The Higgs may decays to other light scalars in Little Higgs models and by understanding the deeper structure of these theories,  it may be possible to link new decays to the structure of collective quartics \cite{Kilian:2004pp,Cheung:2006nk,Reuter:2007gu}.

The classification of collective quartic described in this article only applies to models where the global symmetry group, $G$, is simple.  Many LH models are described by product groups \cite{ArkaniHamed:2001nc,ArkaniHamed:2002pa,ArkaniHamed:2002qx,Chang:2003un,Kaplan:2003uc}.   All of these theories have more pseudo-Goldstone bosons than those considered in this article, but their overall structure may be simpler to realize in ultraviolet completions of Little Higgs models.  Therefore, extending this work to product symmetry groups could be a fruitful and interesting pursuit.
In addition to product groups, the structure of $T$-parity models could be further elucidated from this work \cite{Cheng:2003ju,Cheng:2004yc}

As a final note, this article did not prove whether  it is possible to have $\mathbf{3}_0$ source field without having a dangerous tadpole; however, no such models were found.  The impossibility of such  models was conjectured in \cite{Schmaltz:2008vd} and a small reward was offered. 
Further developments of the techniques in this article may  prove this conjecture.

\section*{Acknowledgments}
We thank Martin Schmaltz, Takemichi Okui and Jesse Thaler for useful discussions.
We thank  Daniele Alves, Kassa Betre, and Eder Izaguirre and especially Mariangela Lisanti for useful comments on the draft.
JGW would like to thank Josephine Suh for early collaboration on this work.
AH  and JGW are supported by the US DOE under contract number DE-AC02-76SF00515 and receive partial support from the Stanford Institute for Theoretical Physics.  
JGW is partially supported by the US DOE's Outstanding Junior Investigator Award.  JGW thanks the Galileo Galilei Institute for their hospitality during the later stages of this work.


\end{document}